\documentclass[aps,prc,twocolumn,showpacs,superscriptaddress,floatfix,nofootinbib]{revtex4}
  
\usepackage{graphicx}
\usepackage{epstopdf}
\usepackage{dcolumn}
\usepackage{amsfonts}
\usepackage{braket}
\usepackage{amssymb,amsmath}
\usepackage{cancel}
\usepackage{comment}
\usepackage{hyperref}

\def\be{\begin{equation}}
\def\ee{\end{equation}}
\def\ba{\begin{eqnarray}}
\def\ea{\end{eqnarray}}
\def\bas{\begin{eqnarray*}}
\def\eas{\end{eqnarray*}}

\begin{document}

\title{Rotational constants of multi-phonon bands in an effective theory for deformed nuclei}

\author{Jialin Zhang} 

\affiliation{Department of Physics and Astronomy, University of
  Tennessee, Knoxville, Tennessee 37996, USA}

\author{T. Papenbrock} 

\affiliation{Department of Physics and Astronomy, University of
  Tennessee, Knoxville, Tennessee 37996, USA}

\affiliation{Physics Division, Oak Ridge National Laboratory, Oak
  Ridge, Tennessee 37831, USA}

\date{\today}

\begin{abstract} 
  We consider deformed nuclei within an effective theory that exploits
  the small ratio between rotational and vibrational excitations. For
  even-even nuclei, the effective theory predicts small changes in the
  rotational constants of bands built on multi-phonon excitations that
  are linear in the number of excited phonons. In $^{166,168}$Er, this
  explains the main variations of the rotational constants of the
  two-phonon $\gamma$ vibrational bands. In $^{232}$Th, the effective
  theory correctly explains the trend that the rotational constants
  decrease with increasing spin of the band head. We also study the
  effective theory for deformed odd nuclei. Here, time-odd terms enter
  the Lagrangian and generate effective magnetic forces that yield the
  high level densities observed in such nuclei.
\end{abstract}


\maketitle

\section{Introduction}

Deformed nuclei exhibit rotational bands as their lowest excitations,
with actinides and rare earth nuclei being the most prominent and best
studied examples~\cite{davidson1981,aprahamian2006}. The
theoretical description and understanding of these nuclei largely
rests on the Bohr Hamiltonian~\cite{bohr_1952}, the collective model
by Bohr and Mottelson~\cite{bohrmottelson_1953,bmbook}, its extension
within the general geometric
models~\cite{Faessler1965,eisenberg,Gneuss1971,hess1980}, and
algebraic models~\cite{arima1975,iachello}. For even-even nuclei the
geometrical models employ rotations and shape parameters as the
relevant degrees of freedom, while algebraic models utilize bosonic
degrees of freedom.  The theoretical approach to odd-mass nuclei is
more cumbersome and is based on coupling the odd nucleon to an
even-even nucleus~\cite{kerman1956,iachello1979}. More microscopic
approaches to deformed nuclei can be based on mean-field
calculations~\cite{aberg1990,frauendorf2001} and shell-model
studies~\cite{vargas2000}.  Being solidly based on fermionic degrees
of freedom, the microscopic models can properly illuminate interesting
phenomena such as, e.g., the effect of pairing on nuclear moments of
inertia~\cite{halbert1993,dracoulis1998,zhang2009,wu2011}.

The collective models are particularly successful in certain symmetry
limits of the Hamiltonian (or for certain choices of the potential
energy) where analytical solutions are available.  Away from these
limits, generalizations of collective models employ expansions of
kinetic and potential terms, or expansions in the number of boson
operators.  Such approaches can be systematic but lack a power
counting, i.e.  higher-order terms in the Hamiltonian are not
guaranteed to yield smaller contributions than low order terms. This
difficulty compounds the adjustment of model
parameters~\cite{hess1980}. Recently, a computationally tractable
approach to the collective model was proposed by Rowe~\cite{Rowe2004},
and applied to the Bohr model~\cite{rowe2009}. Some of the challenges
in the theory of deformed nuclei are described in
Ref.~\cite{matsuyanagi2010}.

An alternative approach to deformed nuclei can be formulated as an
effective theory~\cite{papenbrock2011}. This approach employs similar
degrees of freedom as the Bohr Hamiltonian, and its highlights are the
non-linear realization of rotational symmetry (as a consequence of the
spontaneous symmetry breaking associated with nuclear deformation),
and a power counting. It is thus similar in spirit to other effective
field theories~\cite{Weinberg1990,Weinberg1991} that have been
employed to describe nuclear
interactions~\cite{vankolck1994,bedaque2002,Epelbaum2006,Machleidt2011}, halo
nuclei~\cite{bertulani2002, higa2008} and dilute Fermi
systems~\cite{marini1998,papenbrock1999,hammer2000,Furnstahl2007}.

At next-to-leading order, the effective theory for deformed even-even
nuclei yields spectra that agree (to this order) with those from the
Bohr Hamiltonian, i.e. vibrational states serve as band heads of
rotational bands, with all bands exhibiting the same moment of
inertia~\cite{papenbrock2011}. However, the phenomenology is richer
and more complicated.  Deformed nuclei typically exhibit small
variations in the rotational constants of individual bands, and
accounting for the observation~\cite{davidson1981} that rotational
constants decrease with increasing energy of the band head is a
longstanding problem for the traditional collective models for
well-deformed~\cite{warner1981,Grosse1981,kuyucak1988,caprio2005,Caprio2009,rowe2009,Caprio2011,heyde2011}
and transitional nuclei~\cite{iachello2001,casten2001,pietralla2004}.
To address this problem, we extend the effective theory of deformed
nuclei to next-to-next-to-leading order. 

Another interesting problem concerns deformed odd-mass nuclei. Though
accounting for half of all deformed nuclei, our understanding of them
is much more limited, and the theoretical approach more complicated,
than is the case for even-even nuclei.  Within the collective models
such nuclei are described by coupling a nucleon to an even-even
nucleus~\cite{kerman1956,eisenberg}, or within boson-fermion
models~\cite{iachello1979}. The presence of the odd fermion compounds
the description of odd-mass nuclei considerably.  The question thus
arises whether the odd nucleon really is a degree of freedom that is
relevant at low energies, or to what extent collective vibrations and
rotations alone are sufficient to describe low-energy phenomena of odd
nuclei. In this paper, we will address this question by constructing
the effective theory for deformed odd-mass nuclei at next-to-leading
order.

This paper is organized as follows. Section~\ref{sec:intro} introduces
the effective theory for deformed nuclei. In Sect.~\ref{sec:ee} we
derive the couplings between rotations and vibrations at
next-to-next-to leading order, and compute the resulting spectrum. We
confront theory and data in Section~\ref{sec:compare}.
Section~\ref{sec:odd} extends the effective theory for odd-mass nuclei
to next-to-leading order. Finally, a summary of our results is
presented in Sect.~\ref{sec:conclu}.

\section{Effective theory for deformed nuclei}\label{sec:intro}

An effective theory for deformed nuclei with axial symmetry was
derived in Ref.~\cite{papenbrock2011}. Here we summarize the essential
ingredients of the theory and contrast it to the collective model.

The effective theory is based on quadrupole degrees of freedom
$\phi_\mu(t)$, $\mu=-2,-1, \ldots,2$ because these are sufficient to
reproduce the spins and parities of low-lying states in even-even
nuclei. The reality condition $\phi_{-\mu}=(-1)^\mu\phi_\mu^*$
expresses invariance under time reversal and implies that we deal with
five real degrees of freedom. We assume the spontaneous breaking of
rotational symmetry and a nonzero expectation value
$\langle\phi_0\rangle=v > 0$.  This implies the existence of two
Nambu-Goldstone modes, which may be chosen as the Euler angles
$\alpha(t)$ and $\beta(t)$ that change the orientation of the axially
symmetric nucleus. The three remaining degrees of freedom are chosen
as the complex ``field'' $\phi_2(t)$ and the real ``field''
$\phi_0(t)$.
\begin{equation}
\phi =\begin{pmatrix}
\phi_{2} \\
0\\
\phi_{0}\\
0\\
\phi_{-2} 
\end{pmatrix}\;.
\end{equation}
Thus, the complex ``field'' $\phi_{1}(t)$ is replaced by the two
Nambu-Goldstone bosons. This is consistent with the choice of $\phi_0$
having a nonzero expectation value $v$~\cite{weinbergbook}: An
infinitesimal rotation of the configuration with components
$\phi_\mu=v\delta_{0\mu}$ will generate nonzero components $\phi_{\pm
  1}$. It is convenient to rewrite $\phi_0$ in terms of its vacuum
expectation value $v$ and a small fluctuating part $\varphi_0$ as
\begin{equation}
\phi_0(t)=v+\varphi_0(t) \ .
\end{equation}
We must assume that $|\varphi_0|\ll v $ because of the spontaneous
breaking of rotational symmetry.

Due to the spontaneous symmetry breaking, the rotational symmetry is
realized nonlinearly, and quantities with proper transformation
properties are
\begin{eqnarray}
E_{x}&=&\dot{\alpha}\sin\beta\;,\notag\\ 
E_{y}&=&-\dot{\beta}\;.
\end{eqnarray} 
Under a general rotation by the Euler angles
$(\varphi_1,\varphi_2,\varphi_3)$, the quantities $E_x$ and $E_y$
transform as the $x$ and $y$ components, respectively, of a vector
under a rotation around the $z$ axis by a complicated angle
$\eta(\varphi_1,\varphi_2,\varphi_3,\alpha,\beta)$. The exact
transformation is of no interest here but can be found in
Ref.~\cite{papenbrock2011}.  Thus, the linear combinations
\begin{equation}
E_{\pm}=E_{x} \mp i E_{y}\;
\end{equation}
transform under a rotation as $E_\pm \to e^{\mp i\eta}E_\pm$. 

Likewise, the quadrupole fields transform as $\phi_\mu\to
e^{-i\mu\eta}\phi_\mu$ under a rotation. The covariant derivative
\begin{equation}
D_{t} \equiv \partial _{t} - i E_{z}J_{z} \;,
\end{equation}
with 
\begin{equation}
E_{z}=-\dot{\alpha}\cos\beta   
\end{equation}
is invariant under rotations because $E_z$ transforms as a gauge field.
Here, $J_z$ is the $z$ component of an angular momentum, i.e.
$J_z E_\pm = \pm E_\pm$ and $J_z\phi_\mu = \mu\phi_\mu$.

Due to the nonlinear realization of rotational symmetry, any
Lagrangian that consists of $E_\pm, \phi_{\pm 2}, \phi_0, D_t$ and is
formally invariant under axial (i.e. SO(2)) symmetry is indeed
invariant under full rotational (i.e. SO(3)) symmetry.

For the systematic construction of Lagrangians one needs to establish
a power counting.  We denote the energy scale of rotational
excitations as $\xi$ and that of vibrational excitations as $\Omega$.
One has $\xi \ll \Omega$ with typical values of $\xi\approx 100$~keV
and $\Omega\approx 1$~MeV in rare earth nuclei. For actinides, the
typical values for $\xi$ are smaller by about a factor of two. We also
have to identify a breakdown scale $\Lambda$ of our effective theory.
The complete spectroscopy of low-lying levels in deformed nuclei has
been reported for $^{168}$Er~\cite{davidson1981} and
$^{162}$Dy~\cite{aprahamian2006}. The existence of negative parity
bands in these nuclei (which would require the introduction of
octupole degrees of freedom), and the absence of clear signatures for
multi-phonon vibrations indicates that $\Lambda= \kappa \Omega$ with
$\kappa\approx 2$ or $3$. For the quantities introduced so far the
power counting is
\begin{eqnarray}
\label{power}
E_\pm \sim E_z &\sim& \xi\;,\notag\\
D_{t}\phi_{0}\sim D_{t}\phi_{2} &\sim& \Omega^{1/2}\;,\notag\\
\varphi_{0}\sim\phi_{2} &\sim& \Omega^{-1/2}\;,\notag\\
\phi_{0}\sim v &\sim& \xi^{-1/2}\;.
\end{eqnarray}
This power counting is based on the following rationale: The angles
$\alpha$ and $\beta$ are dimensionless, and a time derivative of these
fields (as in $E_\pm$ and $E_z$) must scale as the low-energy scale
$\xi$. Likewise, a time derivative on the field $\phi$ must scale as
$\Omega$, and the scaling of the fields $\phi_2,\varphi_0$ itself
ensures that the kinetic term $(D_t\phi)^2$ scales as $\Omega$.
Finally, the expectation value $v$ is associated with the spontaneous
symmetry breaking and must thus scale as $\xi^{-1/2}$. In an infinite
system, we would have $\xi\to 0$, correctly implying both the
divergence of the vacuum expectation value $v$ and zero-energy
Nambu-Goldstone modes.

Let us briefly recapitulate the effective theory for deformed nuclei
at next-to-leading order for even-even nuclei~\cite{papenbrock2011}.
At leading order, i.e. at order $\Omega$, we have only vibrations, and
we note that 
\begin{equation}
\label{Tkin}
(D_t\phi_2)(D_t\phi_{-2}) =\dot{\phi}_2\dot{\phi}_{2}^* 
- 4{\rm Im}\left(\dot\phi_2\phi_2^*\right) E_z 
+ 4\phi_2\phi_{2}^*E_z^2
\end{equation}
consists of three terms that are suppressed by subsequent
factors of $\xi/\Omega$ when going from left to right. 

The Langrangian at LO is
\begin{eqnarray}
\label{nlolag}
L_{\rm LO}=\frac{1}{2}\dot\phi_0^2+\dot\phi_2 \dot\phi_{-2} 
-\dfrac{\omega_{0}^2}{2}\varphi_{0}^2-\dfrac{\omega_{2}^2}{4}\phi_{2}\phi_{-2} \ .
\end{eqnarray}
Here, we assume that $\omega_0\sim\omega_2\sim\Omega$. We use
$\phi_2=\varphi_2 e^{i\gamma}$ with real $\varphi_2$ and $\gamma$, and
perform the Legendre transformation
\begin{eqnarray}
p_0&=&{\partial L_{\rm LO} \over \partial \dot\varphi_0}\ ,\nonumber\\
p_2&=&{\partial L_{\rm LO} \over \partial \dot\varphi_2}\ , \nonumber\\
p_\gamma&=&{\partial L_{\rm LO} \over \partial \dot \gamma} \ .\nonumber\\
\end{eqnarray}
The Hamiltonian is
\begin{eqnarray}
H_{\rm LO}=\frac{p_0^2}{2} +{\omega_0^2\over 2}\varphi_0^2 +  
{1\over 4}\left(p_2^2 + {p_\gamma^2\over \varphi_2^2}\right) + {\omega_2^2\over 4}\varphi_2^2 \ , 
\end{eqnarray}
and the spectrum is thus equal to the one of an axially symmetric
harmonic oscillator in three spatial dimensions with energies
\begin{equation}
\label{lospec}
E_{\rm LO}(n_0,n_2,l_2) = \omega_0 (n_0 +1/2)  +{\omega_2\over 2} (2n_2+|l_2|+1) \ .
\end{equation}
With view on the breakdown scale $\Lambda$ of the effective theory, we
limit ourselves to the ground state with quantum numbers
$(n_0,n_2,l_2)=(0,0,0)$, and the two lowest vibrational states with
quantum numbers $(1,0,0)$ and $(0,0,1)$, respectively. The
eigenfunctions are products
\begin{equation}
\Psi_{\rm LO}(\gamma,\varphi_0,\varphi_2)=e^{-il_2\gamma}  \psi_{n_0}(\varphi_0)\chi_{n_2 l_2}(\varphi_2) \ .
\end{equation}
Here, $\psi_{n_0}(\varphi_0)$ is the eigenfunction of the
one-dimensional harmonic oscillator with frequency $\omega_0$, while
$\chi_{n_2 l_2}(\varphi_2)$ is the radial eigenfunction of the
two-dimensional isotropic oscillator with frequency $\omega_2$.

At next-to-leading order, the Nambu-Goldstone modes enter in addition
to higher order corrections in the kinetic energy~(\ref{Tkin}), and
the Lagrangian becomes with
\begin{eqnarray}
\label{eq:nglag}
L_{\rm NLO}&=&L_{\rm LO}+\Delta L_{\rm NLO}\nonumber\\
\Delta L_{\textrm{NLO}}&=&\frac{C_{0}}{2}E_{+}E_{-}- 4{\rm Im}\left(\dot\phi_2\phi_2^*\right)E_z\nonumber\\ 
&=&\frac{C_{0}}{2}\left(\dot{\beta}^2+\dot{\alpha}^2\sin^2\beta\right)+4\varphi_2^2\dot\gamma \dot\alpha\cos\beta\;.
\end{eqnarray}
Here, we assume that $C_0\sim\xi^{-1}$, and the NLO correction is thus
of order $\xi$. Note that we neglected next-to-leading order
corrections (``anharmonicities'') to the vibrational potential. Such
anharmonicities would affect higher-lying vibrational states (which
are at or beyond the breakdown scale $\Lambda$ of the effective
theory) and transition matrix elements (which are not the interest of
this work).  The Hamiltonian at NLO thus becomes
\begin{eqnarray}
H_{\textrm{NLO}}&=&\frac{1}{2}p_{0}^2+\frac{1}{4}p_{2}^2+\frac{p_{\gamma}^2}{4\varphi_{2}^2}+\frac{\omega_{0}^2}{2}\varphi_{0}^2+\frac{\omega_{2}^2}{4}\varphi_{2}^2\notag\\
&+&\frac{1}{2C_{0}}\left( p_{\beta}^2+\frac{1}{\sin^2\beta}(p_{\alpha}-2p_{\gamma}\cos\beta)^2\right)\;.
\label{eq:nonptbnloham}
\end{eqnarray}
The corresponding energy spectrum is
\begin{eqnarray}
E_{\rm NLO}(n_0,n_2,l_2,I)&=&E_{\rm LO}(n_0,n_2,l_2)\nonumber\\
&+&{I(I+1)-(2l_{2})^2\over 2C_0}
\;, \label{eq:nloenergy}
\end{eqnarray}
and the eigenfunctions are 
\begin{eqnarray}
\lefteqn{\Psi_{\rm NLO}(\alpha,\beta,\gamma,\varphi_0,\varphi_2)=}\nonumber\\
&&e^{-im\alpha} d_{m,2l_2}^I(\beta)  \Psi_{\rm LO}(\gamma,\varphi_0,\varphi_2) \ .
\end{eqnarray}
Here, $I \ge |2l_2|$ denotes the angular momentum, and $m$ the
angular-momentum projection with $-I\le m\le I$. The eigenfunction
$d_{\mu,\nu}^I(\beta)$ is part of the Wigner $D$ function
$D_{\mu,\nu}^I(\alpha,\beta,\gamma)=e^{-i\mu\alpha}d_{\mu,\nu}^I(\beta)e^{-i\nu\gamma}$.
Thus, we can rewrite
\begin{eqnarray}
\label{eigenf}
\lefteqn{\Psi_{\rm NLO}(\alpha,\beta,\gamma,\varphi_0,\varphi_2)=}\nonumber\\
&& D_{m,2l_2}^I(\alpha,\beta,\gamma)\psi_{n_0}(\varphi_0)\chi_{n_2 l_2}(\varphi_2)\ .
\end{eqnarray}

The spectrum~(\ref{eq:nloenergy}) consists of rotational bands
(labeled by the angular momentum $I$) on top of the vibrational band
heads (labeled by the quantum numbers $n_0, n_2, l_2$). Note that the
moment of inertia $C_0$ is identical for every rotational band. 

Let us also compare the effective theory with the Bohr
model.  Recall that the Bohr model starts from five quadrupole degrees
of freedom, and a transformation to the body-fixed coordinate system
yields three Euler angles and two shape parameters (usually denoted as
$\beta$ and $\gamma$). The $\beta$ degree of freedom corresponds to
axially symmetric oscillations around the static deformation while
$\gamma$ accounts for triaxial deformations. In the Bohr Hamiltonian,
the vibrational and rotational degrees of freedom are coupled via the
moment of inertia, while the effective theory is less constrained.
Bohr's $\beta$ degree of freedom corresponds to $\varphi_0$ in the
effective theory. One can combine Bohr's $\gamma$ degree of freedom
and Bohr's rotational angle $\psi$ to a two-dimensional harmonic
oscillator~\cite{bohrmottelson1982}. In this combination, these two
degrees of freedom correspond to the complex $\phi_2$ (or $\varphi_2$
and $\gamma$) in the effective theory.  Let us introduce
\begin{equation}
\label{K}
K\equiv 2l_2
\end{equation}
for the third quantum number of the axially symmetric rotor.  With
this notation, the effective theory at NLO is in agreement with the
spectra and wave functions obtained for the collective model ({\it
  cf.} chapter 6 of Ref.~\cite{eisenberg}). This agreement is
expected.

\section{Even-even nuclei at next-to-next-to-leading order}\label{sec:ee}

At NNLO we have to include terms of the size $~\xi^2/\Omega$. As
before, we focus on the terms that couple rotations and vibrations.
This is perhaps one of the main differences between the collective
model and the effective theory. In the former, most authors have
restricted themselves to study higher order corrections to the
vibrational potential. This is presumably due to the difficulty to
write down (and to work with) higher order corrections to the kinetic
terms. In the effective theory, this task is straightforward and
yields~\cite{papenbrock2011}
\begin{eqnarray}
L_{\textrm{NNLO}}&=&L_{\textrm{NLO}}+4\phi_2\phi_2^*E_z^2+ \Delta L_{\textrm{NNLO}}\;, \label{eq: nnlolag}\\
\Delta L_{\textrm{NNLO}}&=&D_{0}(E_{+}E_{-})\varphi_{0}^2+F_{0}(E_{+}E_{-})\dot{\varphi_{0}}^2\notag\\
&+&D_{2}(E_{+}E_{-})|\phi_{2}|^2+F_{2}(E_{+}E_{-})|D_{t}\phi_{2}|^2\notag\\
&+&D_{1}\varphi_{0}(\phi_{2}E_{-}^2+\phi_{-2}E_{+}^2)\notag\\
&+&F_{1}\dot{\varphi_{0}}(E_{+}^2D_{t}\phi_{-2}+E_{-}^2D_{t}\phi_{+2})\;.  
\label{eq:nnlodlag}
\end{eqnarray}
Here, $\Delta L_{\textrm{NNLO}}$ denotes the rotation-vibration
interaction at NNLO. Each term in $\Delta L_{\textrm{NNLO}}$ has the
order of magnitude $\mathcal{O}(\xi^2/\Omega)$, making the
undetermined coefficients scale as
\begin{eqnarray}\label{nnloparascaling}
D_{0}&\sim& D_{1}\sim D_{2} \sim {\cal O}(1)\notag\;,\\
F_{0}&\sim& F_{1} \sim F_{2} \sim \Omega^{-2}\;.
\end{eqnarray}
The correctness of these scaling relations should be validated by
fitting the derived spectrum to the experimental level schemes.

The Lagrangian $L_{\textrm{NNLO}}$ expanded in terms of the polar
coordinates $\varphi_2$ and $\gamma$ and the Euler angles $\alpha$ and
$\beta$ is
\begin{eqnarray}
L_{\textrm{NNLO}}&=&\frac{1}{2}\dot{\varphi_{0}}^2+\dot{\varphi_{2}}^2+\varphi_{2}^2\dot{\gamma}^2-\frac{\omega_{0}^2}{2}\varphi_{0}^2-\frac{\omega_{2}^2}{4}\varphi_{2}^2\nonumber\\
&+&4\varphi_{2}^2\left(\dot{\gamma}+\dot{\alpha}\cos\beta\right)\dot{\alpha}\cos\beta\nonumber\\
&+&\frac{C_{0}}{2}\left(\dot{\beta}^2 +\dot{\alpha}^2\sin^2\beta\right)
+\Delta L_{\textrm{NNLO}}\;, \label{eq:nnlolag}
\end{eqnarray}
with
\begin{eqnarray}
\Delta L_{\textrm{NNLO}}&=&\left(\dot{\beta}^2+\dot{\alpha}^2\sin^2\beta\right)\Big[D_{0}\varphi_{0}^2+F_{0}\dot{\varphi_{0}}^2 \nonumber\\
&&+D_{2}\varphi_{2}^2+F_{2}\left(\dot{\varphi_{2}}^2+\varphi_{2}^2\dot{\gamma}^2\right)\Big]\nonumber\\
&+&2\left(\dot{\alpha}^2\sin^2\beta-\dot{\beta}^2\right)
\Big[D_{1}\varphi_{0}\varphi_{2}\cos\gamma \nonumber\\
&&+ F_{1}\dot{\varphi_{0}}(\dot{\varphi_{2}}\cos\gamma-\varphi_{2}\dot{\gamma}\sin\gamma)\Big]\nonumber\\
&+&4\dot{\alpha}\dot{\beta}\sin\beta\Big[D_1\varphi_0\varphi_{2}\sin\gamma\nonumber\\
&&+F_1\dot{\varphi}_0(\dot{\varphi_{2}}\sin\gamma+\varphi_{2}\dot{\gamma}\cos\gamma)\Big]\;.
\label{eq:deltaL}
\end{eqnarray}
It is difficult to perform the Legendre transformation rigorously on
$L_{\textrm{NNLO}}$, because $\Delta L_{\textrm{NNLO}}$ admixes the
Nambu-Goldstone modes and quadrupole fields and the velocity-momentum
inversions always involve quadratic terms. Fortunately, we do not need
the perform the Legendre transformation of the
Lagrangian~(\ref{eq:nnlolag}) exactly but rather can employ
perturbation theory for this task.

For this purpose we follow Fukuda and
coworkers~\cite{fukuda1988} who applied perturbative
Legendre transformations to several physics
problems~\cite{Inagaki1992,fukuda1994}.  Fukuda's
inversion method expands the generalized velocities perturbatively
order by order in the small quantity $\xi / \Omega$. For instance,
$\dot{\varphi_0}$ is expanded as
\begin{equation}
\dot{\varphi_0}=\dot{\varphi_0}^{(0)}+\dot{\varphi_0}^{(1)}+\dot{\varphi_0}^{(2)}+\ldots\;.
\end{equation}
Here, $\dot{\varphi_0}^{(0)}$ has the same order of magnitude as
$\dot{\varphi_0}$ and is of leading order. Higher-order corrections
scale as
\begin{equation}
\dot{\varphi_0}^{(i+1)}\sim \dot{\varphi_0}^{(i)}\frac{\xi}{\Omega}\;.  
\end{equation}
The key step consists of assuming the generalized momenta to be of
leading order (and with no further corrections).  Thus, the
leading-order relation between the momenta and velocities of the
Lagrangian~(\ref{eq:nnlolag}) is
\begin{eqnarray}
\label{eq:loperturb}
p_{0}&=&\dot{\varphi_{0}}^{(0)}\;,\notag\\
p_{2}&=&2\dot{\varphi_{2}}^{(0)}\;,\notag\\
p_{\gamma}&=&2\varphi_{2}^2\dot{\gamma}^{(0)}\;,\notag\\
p_{\alpha}&=&C_{0}\dot{\alpha}^{(0)}\sin^2\beta+4\varphi_{2}^2\dot{\gamma}^{(0)}\cos\beta\;,\notag\\
p_{\beta}&=&C_{0}\dot{\beta}^{(0)}\;.
\end{eqnarray} 
It is straightforward to invert these equations. The higher-order
corrections of the velocities now fulfill homogeneous equations (as
the momenta consist only of leading-order terms), and can be solved
perturbatively to the desired order.  In what follows, we only present
the result of the Legendre transformation of the Lagrangian
Eq.~(\ref{eq:nnlolag}) using the Fukuda's inversion method, and refer
the reader to Ref.~\cite{fukuda1994} for more details.

The Legendre transformation yields the Hamiltonian 
\begin{equation}\label{eq:finalnnloham}
H_{\textrm{NNLO}}=H_{\textrm{NLO}}-\Delta L_{\textrm{NNLO}}^{(0)}\;.
\end{equation}
Here $H_{\textrm{NLO}}$ is the NLO Hamiltonian given in
Eq.~(\ref{eq:nonptbnloham}), and the term $\Delta
L_{\textrm{NNLO}}^{(0)}$ is from Eq.~(\ref{eq:deltaL}) with all
leading-order velocities re-expressed in terms of momenta
(\ref{eq:loperturb}) and all higher-order velocities dropped in this
term.

The eigenvalues of $H_{\textrm{NLO}}$ are given in
Eq.~(\ref{eq:nloenergy}) and the small contribution of $\Delta
L_{\textrm{NNLO}}^{(0)}$ to the spectrum can be worked out in
perturbation theory by computing the expectation value of $\Delta
L_{\rm NNLO}^{(0)}$ in the eigenstates~(\ref{eigenf}) of the
Hamiltonian~(\ref{eq:nonptbnloham}). For computation of the
expectation value $\langle
(\dot{\alpha}^{(0)})^2\sin^2\beta+(\dot{\beta}^{(0)})^2\rangle$ we
note that
\begin{eqnarray}
\lefteqn{\left(\dot{\alpha}^{(0)}\right)^2\sin^2\beta+(\dot{\beta}^{(0)})^2}\nonumber\\
&=&\frac{1}{C_{0}^2}\left(\frac{1}{\sin^2\beta}(p_{\alpha}-2p_{\gamma}\cos\beta)^2+p_{\beta}^2\right)\nonumber\\
&=&\frac{1}{C_{0}^2}\left(I(I+1)-(2l_{2})^2\right) \ .
\end{eqnarray}
For the expectation values involving the quadrupole vibrations we have
\begin{eqnarray}
\langle \varphi_{0}\rangle &=&\langle \dot{\varphi_{0}}^{(0)}\rangle=0\;,\nonumber\\
\langle \varphi_{0}^2\rangle &=& \frac{1}{\omega_{0}}\left(n_{0}+\frac{1}{2}\right)\;,\notag\\
\langle (\dot{\varphi_{0}}^{(0)})^2\rangle &=& \omega_{0}\left(n_{0}+\frac{1}{2}\right)\;,\notag\\
\langle\varphi_{2}^2\rangle&=&\frac{1}{\omega_{2}}(2n_{2}+|l_{2}|+1)\;,\notag\\
\langle(\dot{\varphi_{2}}^{(0)})^2+\varphi_{2}^2(\dot{\gamma}^{(0)})^2\rangle&=&\frac{\omega_2}{4}(2n_{2}+|l_{2}|+1)\;.
\end{eqnarray}
Hence, we find 
\begin{eqnarray}
\langle\Delta L_{\textrm{NNLO}}^{(0)}\rangle &=&\frac{I(I+1)-(2l_{2})^2}{2C_{0}}
\bigg[\left(n_{0}+\frac{1}{2}\right)R\nonumber\\
&+&(2n_{2}+|l_{2}|+1)S\bigg]\;.
\label{shift}
\end{eqnarray}
Here, we used the shorthands
\begin{eqnarray}
R&\equiv&\frac{2}{C_0}\left(\frac{D_{0}}{\omega_{0}} +F_0\omega_{0}\right)\;,\notag\\
S&\equiv&\frac{2}{C_0}\left(\frac{D_{2}}{\omega _{2}}+\frac{1}{4}F_2\omega _2\right)\;.\label{eq:redefconsts}
\end{eqnarray}
Thus, the next-to-next-to-leading order correction to the
energies~(\ref{eq:nloenergy}) is the small shift~(\ref{shift}) of
order ${\cal O}(\xi^2/\Omega)$. This shift yields corrections to the
moments of inertia of the different rotational bands and depends on
the quantum numbers $(n_0,n_2,l_2)$ of the band head. In particular,
the moment of inertia of the $\beta$ band depends on $R$
while that of the $\gamma$ band depends on $S$. Thus, the rotational
bands of multi-phonon excitations have rotational constants
\begin{equation}
\label{rotconst1}
A_{\rm theo} = {1-\left(n_{0}+\frac{1}{2}\right)R - (2n_{2}+|l_{2}|+1)S\over 2C_0} \ .
\end{equation}
In practice it is useful to rewrite this expression as
\begin{equation}
\label{rotconst}
A_{\rm theo}= A_{\rm g.s.} - a_\beta n_{0} - a_\gamma (2n_{2}+|K|/2)\ .
\end{equation}
Here, $A_{\rm g.s.}$ is the rotational constant of the ground-state
band, and $a_\beta$ and $a_\gamma$ denote the small corrections for
bands built on multi-phonon excitations. We used the
relation~(\ref{K}). As usual, $A_{\rm theo} [I(I+1)-K^2]$ describes
the energy levels of rotational bands. Note that the change in the
rotational constants is linear in the number of excited phonons.  This
is one of the main result of this paper.  The small correction to the
moment of inertia depends on the parameters $a_\beta$ and $a_\gamma$
(or $R$ and $S$), and can be determined by fit to data.  Note that the
terms in Eq.~(\ref{eq:deltaL}) proportional to $D_1$ and $F_1$ do not
affect the spectrum at next-to-next-to leading order because of the
zero expectation values of the position $\varphi_0$ and velocity
$\dot{\varphi}_0$ of the harmonic oscillator.  These terms will affect
wave functions at the considered order and spectra at the next higher
order.

\section{Comparison between theory and data}
\label{sec:compare}

Let us confront our predictions with data. The effective theory we
derived allows us to describe small deviations in the moment of
inertia of the $\beta$ band and the $K=2$ $\gamma$ band by a fit of
$R$ and $S$, respectively. The theory is thus sufficiently flexible to
accommodate the small differences between the observed rotational
constants for the ground-state band and the $\beta$ and $\gamma$ bands
of a deformed nucleus. This overcomes a deficiency of the collective
models, see e.g.
Refs.~\cite{warner1981,kuyucak1988,casten2001,Caprio2009,Caprio2011}.
The Table~I in Ref.~\cite{heyde2011} shows that $a_\beta$ is positive
for most deformed nuclei. Once the low-energy constants $C_0$, $R$ and
$S$ (or $A_{\rm g.s.}$, $a_\beta$ and $a_\gamma$) are determined from
the ground-state, the $\beta$, and the $\gamma$ bands, the effective
theory predicts that the difference between the rotational constants
of multi-phonon vibrations and the ground-state band depends linearly
on the number of excited phonons.  There are only a few candidates for
two-phonon excitations in deformed nuclei, see
Refs.~\cite{Sood1991,Sood1992} for a summary of the status of the
field in the early 1990s. Due to experimental advances, there is now
robust evidence for two-phonon $\gamma$-vibrational excitations in
$^{168}$Er\cite{Boerner1991,Oshima1995,Haertlein1998},
$^{166}$Er~\cite{Fahlander1996,Garrett1997}, and
$^{232}$Th~\cite{Korten1993,martin2000}. For earlier theoretical
discussions on multi-phonon states in $^{168}$Er, we refer the reader
to
Refs.~\cite{warner1981,bohrmottelson1982,matsuo1984,matsuo1985,Piepenbring1988}.

Table~\ref{tab1} summarizes our results for $^{168,166}$Er and
$^{232}$Th, respectively. The Table shows the excitation energy $E$ of
the band head, its spin $K$, and the rotational constant $A$. The
latter was deermined by computing the first level spacing of the
respective rotational bands according to the formula $A[I(I+1)-K^2]$.
For each nucleus, the theoretical rotational constants $A_{\rm theo}$
are determined by adjusting the low-energy constants $A_{\rm g.s.}$
and $a_\gamma$ of Eq.~(\ref{rotconst}) to the rotational constants of
the ground-state band and the $\gamma$ band.  This yields
$a_\gamma=0.84$~keV, $a_\gamma=1.18$~keV, and $a_\gamma=0.85$~keV for
$^{168}$Er, $^{166}$Er, and $^{232}$Th, respectively. These
corrections are much smaller (i.e. by about a factor $\xi/\Omega$) 
than the rotational constant $A_{\rm
  g.s.}=13.17$~keV, $A_{\rm g.s.}=13.43$~keV, and $A_{\rm
  theo}=8.23$~keV of the respective ground-state bands.  For $K=4$,
$A_{\rm theo}$ is a prediction.  These predictions are in good
quantitative agreement with data for $^{168}$Er and in
semi-quantitative agreement with the data for $^{166}$Er and
$^{232}$Th. More precisely, for $^{168}$Er, the difference between
data and theory is about 10\% of $a_\gamma$ and thus consistent with
neglected higher-order corrections [which are of order ${\cal
  O}(\xi/\Omega)$]. For $^{166}$Er, the difference between data and
theory is about 43\% of $a_\gamma$. This difference is probably at the
limit of what one expects from estimates within the effective theory.
For $^{232}$Th, the difference between data and theory is about 87\%
of $a_\gamma$ and clearly larger than expected.  Here, the effective
theory only describes correctly the trend that the rotational
constants decrease with increasing spin $K$ of the band head.

\begin{table}[hbt]
\begin{tabular}{|c|ccc|ccc|ccc|}\hline
    & \multicolumn{3}{c|}{$^{168}$Er}&\multicolumn{3}{c|}{$^{166}$Er}&\multicolumn{3}{c|}{$^{232}$Th}
\\\hline
$E$           & 0     &  821 & 2056 & 0     &  786 & 2028 & 0     & 785  & 1414 \\ 
$K$           & 0     & 2    & 4    & 0     & 2    & 4    & 0     & 2    & 4    \\ 
$A$           & 13.17 & 12.33& 11.37& 13.43 & 12.25& 10.56& 8.23  & 7.38 & 7.27 \\\hline
$A_{\rm theo}$& 13.17& 12.33& 11.49&13.43&12.25& 11.07& 8.23  & 7.38 & 6.53 \\\hline
\end{tabular}
\caption{Experimental excitation energies $E$ (in keV) and  spins $K$ of $\gamma$ 
  vibrational band heads in $^{168,166}$Er and $^{232}$Th. The rotational constants $A$ (in keV) are deduced from the first level spacing of the rotational band. In the theoretical description, the $\gamma$ vibrational states have quantum numbers $n_0=0=n_2$, and $l_2=K/2$. The theoretical result $A_{\rm theo}$ (in keV) for the rotational constant is determined by fit to the $K=0$ and $K=2$ bands and is a prediction for the $K=4$ states.} 
\label{tab1}
\end{table}

Note that -- at the considered order in the effective theory -- the
variation in the rotational constants is not affected by the omission
of next-to-next-to-leading order corrections in the potential of the
vibrational degrees of freedom $(\varphi_0,\varphi_2)$. Those
corrections introduce anharmonicities in the vibrational spectrum
(i.e. the energies of the band heads), but they do not influence the
moments of inertia.  Note also, that the effective theory -- at the
here considered order -- yields the rotational bands of the rigid
rotor (which are proportional to $I(I+1)-K^2$). At the next higher order,
i.e. at order ${\cal }(\xi^3/\Omega^2)$, corrections proportional to
$[I(I+1)-K^2]^2$ enter~\cite{papenbrock2011}.

\section{Odd-mass nuclei at next-to-leading order}\label{sec:odd}
Odd-mass nuclei have half-integer spins in their ground states. We
want to describe these nuclei in terms of vibrations and rotations
alone. The elimination of the odd nucleon as an active degree of
freedom leads to an important change in the symmetry properties of the
Lagrangian for the rotations and vibrations. Due to the finite
ground-state spin, the Lagrangians of odd-mass nuclei are not
invariant under time reversal, and terms that are odd under time
reversal need to be included into the description. In
Ref.~\cite{papenbrock2011}, the effective theory for the
Nambu-Goldstone modes of odd-mass nuclei was considered at leading
order. Here, we go one step further and include the vibrational
degrees of freedom and consider the effective theory for deformed
odd-mass nuclei at next-to-leading order.

Let us start with the vibrational degrees of freedom.  The time-odd
and rotationally invariant terms $\phi_{0}D_{t}\phi_{0}$,
$\phi_{2}D_{t}\phi_{-2}$ and its complex conjugate enter as additional
building blocks of the Lagrangian.  Instead of decomposing $\phi_{2}$ in
the polar coordinates as in even-even nuclei, we here decompose it in the
Cartesian coordinates (mostly for its simplicity in gauge
transformation which we will see later)
\begin{eqnarray}
\phi_{2}&=&x+iy\;.
\end{eqnarray}
Hence,
\begin{eqnarray}
\phi_{2}D_{t}\phi_{-2}&=&x\dot{x}+y\dot{y}-i(x\dot{y}-y\dot{x})+2iE_{z}(x^2+y^2)\;,\notag\\
\phi_{0}D_{t}\phi_{0}&=&\phi_{0}\dot{\phi_{0}}=\frac{1}{2}\partial_{t}\left(\phi_{0}^2\right)\;.\label{eq:expansion}
\end{eqnarray}
The power counting Eq.~(\ref{power}) yields the scaling
\begin{equation}
\phi_{2}D_{t}\phi_{-2}\sim\phi_{-2}D_{t}\phi_{2}\sim\phi_{0}D_{t}\phi_{0}\sim {\cal O}(1)\;.
\end{equation}
All leading-order terms of the Lagrangian of even-even nuclei
Eq.~(\ref{nlolag}) also enter for odd-mass nuclei.  The leading order
Lagrangian for odd-mass nuclei thus becomes
\begin{eqnarray}
L^{\rm (odd)}_{\textrm{LO}}&=&(D_{t}\phi_{2})(D_{t}\phi_{-2})+\frac{1}{2}\dot{\varphi_{0}}^2+\frac{A}{2}\partial_{t}(\phi_{0}^2)\notag\\
&+&\frac{\tilde{A}}{2}\left(\phi_{2}D_{t}\phi_{-2}+ \phi_{-2}D_{t}\phi_{2}\right)\notag\\
&+&\frac{iB}{2}\left(\phi_{2}D_{t}\phi_{-2}-\phi_{-2}D_{t}\phi_{2}\right)\;.
\end{eqnarray}
Here the parameters $B$, $\widetilde{A}$ and $A$ scale as
\begin{equation}
B\sim\widetilde{A}\sim A\sim\Omega\;.
\end{equation} 
Note that $\phi_{2}D_{t}\phi_{-2}$ and $\phi_{-2}D_{t}\phi_{2}$ are
complex conjugate to each other, so they appear as linear combinations
to yield real values. The terms proportional to $A$ and $\tilde{A}$
are total time derivatives and can thus be dropped from the
Lagrangian. However, it is instructive to keep them for a moment, and
we will soon eliminate them by a gauge transformation.  We employ
Eq.~(\ref{eq:expansion}) and find in leading order
\begin{eqnarray}
\label{oddlaglo}
L^{\rm (odd)}_{\rm LO} &=& \dot{x}^2+\dot{y}^2+\frac{1}{2}\dot{\varphi_{0}}^2
+B(x\dot{y}-y\dot{x})\nonumber\\
&+&\frac{A}{2}\partial_{t}\left(\phi_{0}^2\right)+ {\tilde{A}\over 2}\partial_t\left(x^2+y^2\right)\;.
\end{eqnarray}
Clearly, the nontrivial part of the Lagrangian describes a particle in
three dimensions in a constant magnetic field with strength
proportional to $B$. A Legendre transformation yields the Hamiltonian
\begin{eqnarray}
H_{\textrm{LO}}^{\rm (odd)}&=&\frac{1}{2}(p_{0}-A\phi_{0})^2+\frac{1}{4}\left(p_{x}-\tilde{A}x+By\right)^2\nonumber\\
&+&\frac{1}{4}\left(p_{y}-\tilde{A}y-Bx\right)^2 \ .
\end{eqnarray}
Let us employ a gauge transformation with the phase function
\begin{equation}
\lambda(x, y, \phi_{0})=\frac{\tilde{A}}{2}(x^2+y^2)+\frac{A}{2}\phi_{0}^2\;,
\end{equation}
and gradient
\begin{equation}
\vec{\nabla}\lambda=(\tilde{A}x, \tilde{A}y, A\phi_{0})
\end{equation}
to gauge away the trivial terms proportional to $A$ and $\tilde{A}$.
This yields
\begin{equation}\label{eq:gaugedham}
H_{\textrm{LO}}^{\rm (odd)}=\frac{1}{2}p_{0}^2+\frac{1}{4}(p_{x}+By)^2+\frac{1}{4}(p_{y}-Bx)^2\;.
\end{equation}
At leading order, we thus have free motion in the direction of
$\varphi_0$ and quantized Landau levels in the $xy$ plane.

At next-to-leading order, the Langrangian is
\begin{eqnarray}
L^{\rm (odd)}_{\textrm{NLO}}&=&L^{\rm (odd)}_{\textrm{LO}}+\frac{C_{0}}{2}E_+ E_- +qE_z\nonumber\\
&=&\frac{1}{2}\dot{\varphi_{0}}^2+\dot{x}^2+\dot{y}^2
+B(x\dot{y}-y\dot{x})\nonumber\\
&+&\frac{C_{0}}{2}\left(\dot{\alpha}^2\sin^2\beta+\dot{\beta}^2\right)\nonumber\\
&-&\left[q-4(x\dot{y}-y\dot{x})\right]\dot{\alpha}\cos\beta\;.\label{eq:oonlolag}
\end{eqnarray}
Here, we have dropped the irrelevant terms proportional to $A$ and
$\tilde{A}$ in $L_{\rm LO}^{\rm odd}$.  We identify again the
Lagrangian of a particle on the sphere and note that the term $q E_z =
-q\dot{\alpha}\cos\beta$ is technically a Wess-Zumino term. Under
rotations, this term remains invariant up to a total derivative, and
the parameter $q$ is related to the ground-state spin
~\cite{papenbrock2011}.  The coupling between rotations and vibrations
in the Lagrangian~(\ref{eq:oonlolag}) stems from the covariant
derivative that appears in the leading-order
Lagrangian~(\ref{oddlaglo}), and higher-order terms have been
neglected.

Let us discuss the coupling of the nuclear spin to the vibrations and
rotations which is due to the time-odd terms in the Lagrangian.  The
coupling of the ground-state spin to the Euler angles can be viewed as
a particle on the sphere coupled to a magnetic monopole with charge
$2q$~\cite{wu1976}.  Technically, the vibrations couple to the
ground-state spin via an effective magnetic field $B$ that is
generated by the ground-state spin.  Note that our approach takes the
spin of the ground state as a static quantity and not as a degree of
freedom. This is an approximation that we expect to be valid only for
sizeable spins and low energies. At higher energies, or for small
ground-state spins, the spin is a dynamical quantity and only the
total spin, i.e.  the sum of ground-state spin and the spin $I$
associated with the Euler angles is conserved. Our approach excludes
terms such as the ``Coriolis coupling''~\cite{kerman1956} from the
Langrangian, and it is well known that this coupling has an important,
i.e. leading order, contribution for ground-states (or band heads)
with spin $1/2$~\cite{bmbook}.

At this point, we add a leading-order harmonic potential
\begin{equation}
V_{\rm LO} = {\omega_0^2\over 2}\varphi_0^2 
\end{equation}
in the $\varphi_0$ vibrational degree of freedom (the magnetic field
$B$ is the leading-order contribution to the $\phi_2$ degrees of
freedom), and perform the Legendre transformation to obtain the
Hamiltonian. One finds
\begin{eqnarray}\label{vbhamoo}
H_{\textrm{NLO}}^{\rm (odd)}&=&
{1\over 2C_0} \left[p_\beta^2 + {1\over \sin^2\beta}\left(p_\alpha +(q-2l_2)\cos\beta\right)^2 \right]\nonumber\\
&+&\frac{1}{4}\left(p_{x}^2+p_{y}^2\right)+{B^2\over 4}\left(x^2+y^2\right)-\frac{B}{2}l_2\notag\\
&+& \frac{1}{2}p_{0}^2+{\omega_{0}^2\over 2}\varphi_{0}^2\;.
\end{eqnarray}
Note that $l_2=(xp_{y}-yp_{x})$ is an angular momentum. In the
$\varphi_0$ degree of freedom we have a harmonic oscillation.  Upon
quantization, one finds the usual levels of the one-dimensional
harmonic oscillator.  The $\phi_2=x+iy$ degrees of freedom corresponds
to a charged particle moving in a plane perpendicular to a strong
magnetic field. This yields Landau levels upon quantization. On top of
each of these ``vibrational'' states, one finds a rotational band due
to the Euler angles.

The spectrum of the Hamiltonian for odd-mass nuclei at next-to-leading
order thus is
\begin{eqnarray}\label{finitespinenergy}
E_{\textrm{NLO}}^{\rm (odd)}&=&\omega_{0}\left(n_{0}+\frac{1}{2}\right)
+\frac{|B|}{2}(2n_{2}+|l_{2}|+1)\notag\\
&-&\frac{B}{2}l_{2}+\frac{1}{2C_{0}}\left[ I(I+1)-(q-2l_{2})^2\right]\;.
\end{eqnarray}
The quantum numbers are $n_0=0,1,2,\ldots$ for the harmonic
oscillation of $\varphi_0$, $n_2=0,1,2,\ldots$, $l_2=0, \pm 1, \pm 2,
\ldots$, from the Landau levels, and $I=|q-2l_2|,
|q-2l_2|+1,|q-2l_2|+2,\ldots$ for the rotational bands. The
eigenfunctions are essentially as in Eq.~(\ref{eigenf}) for the
even-even nuclei, but with modification of the indices of the Wigner
$D$ function (and again rewriting $\phi_2=x+iy=\varphi_2 e^{i\gamma}$).

Thus, the spectrum exhibits a large level density close to the ground
state, in qualitative agreement with experimental observations for
odd-mass nuclei. The large degeneracy of the lowest Landau level is
split by the $l_2$-dependent shift of the band head. Next-to-leading
order corrections to the vibrational potential (that we neglected for
convenience) would further modify this picture. Note that $q$ must be
a positive or negative half integer, and the ground state with spin
$|q-2l_2|$ is obtained for the value of $l_2$ that minimizes
$|q-2l_2|$ for the fixed $q$. For negative values of $q$ (and positive
values of $B$), this is achieved for $l_2=0$ in the lowest landau
level, and the spin of the ground state is $|q|$. For positive values
of $q$ (again assuming positive $B$), the ground state has spin 1/2,
and $l_2$ is such that $|q-2l_2|=1/2$. We repeat that the
the effective theory derived in this Section is not valid for band
heads with spin 1/2 because the assumption of a static spin is only
warranted for sizeable spins.  

Thus, the effective theory for odd nuclei is quite similar to the
effective theory for even-even nuclei. Both theories predict a number
of low-lying band heads that are collective vibrations. The comparison
with experimental spectra shows that considerable anharmonicities are
required in practice, i.e. next-to-leading order corrections to the
vibrational Lagrangian must be significant. Within the effective
theory, the higher level density in odd deformed nuclei arises due to
magnetic effects and Landau-level physics.

It would of course be interesting to consider the spin as a dynamic
degree of freedom, and to drive the effective theory for odd-mass
nuclei also to next-to-next-to-leading order. However, many more
time-odd terms contribute, and many new parameters will appear, and
this makes the description of spectra less challenging. Instead, it
might be more interesting to couple electromagnetic fields to the
effective theory and confront low-order results with the considerable
amount of available data.

Note finally that the assumption of a static ground-state spin is
probably not valid for odd-odd nuclei due to the weak coupling between
the odd proton and neutron. Thus, one cannot simply let $q$ assume
integer values and apply the theory derived in this Section to odd-odd
nuclei.

\section{Conclusion}\label{sec:conclu}
In summary, we computed higher-order corrections in the effective
theory for deformed nuclei, and focused particularly on the kinetic
terms that couple rotations and vibrations.  In even-even nuclei, the
next-to-next-to-leading order corrections yield small corrections to
the moments of inertia that are linear in the number of excited
phonons.  When applied to $^{166,168}$Er, the effective theory largely
explains the observed variations of the rotational constants of the
two-phonon $\gamma$ vibrations.  In $^{232}$Th, the theory explains
the trend that rotational constants decrease with increasing spin of
the band head. For odd nuclei, the effective theory at next-to-leading
order includes time-odd terms in the Lagrangian.  This approach
introduces effective magnetic fields into the Hamiltonian and
qualitatively explains observed features such as the high level
densities.

\begin{acknowledgments}
  The authors thank M. Caprio, W. Nazarewicz, and N. Pietralla for
  discussions. This work has been supported by the U.S.  Department of
  Energy under grant Nos. DE-FG02-96ER40963 (University of Tennessee)
  and DE-AC05-00OR22725 with UT-Battelle, LLC (Oak Ridge National
  Laboratory).
\end{acknowledgments}

\onecolumngrid
\bibliographystyle{apsrev}
\bibliography{ref}

\begin{thebibliography}{64}
\expandafter\ifx\csname natexlab\endcsname\relax\def\natexlab#1{#1}\fi
\expandafter\ifx\csname bibnamefont\endcsname\relax
  \def\bibnamefont#1{#1}\fi
\expandafter\ifx\csname bibfnamefont\endcsname\relax
  \def\bibfnamefont#1{#1}\fi
\expandafter\ifx\csname citenamefont\endcsname\relax
  \def\citenamefont#1{#1}\fi
\expandafter\ifx\csname url\endcsname\relax
  \def\url#1{\texttt{#1}}\fi
\expandafter\ifx\csname urlprefix\endcsname\relax\def\urlprefix{URL }\fi
\providecommand{\bibinfo}[2]{#2}
\providecommand{\eprint}[2][]{\url{#2}}

\bibitem[{\citenamefont{Davidson et~al.}(1981)\citenamefont{Davidson, Warner,
  Casten, Schreckenbach, B{\"o}rner, Simic, Stojanovic, Bogdanovic, Koicki,
  Gelletly et~al.}}]{davidson1981}
\bibinfo{author}{\bibfnamefont{W.~F.} \bibnamefont{Davidson}},
  \bibinfo{author}{\bibfnamefont{D.~D.} \bibnamefont{Warner}},
  \bibinfo{author}{\bibfnamefont{R.~F.} \bibnamefont{Casten}},
  \bibinfo{author}{\bibfnamefont{K.}~\bibnamefont{Schreckenbach}},
  \bibinfo{author}{\bibfnamefont{H.~G.} \bibnamefont{B{\"o}rner}},
  \bibinfo{author}{\bibfnamefont{J.}~\bibnamefont{Simic}},
  \bibinfo{author}{\bibfnamefont{M.}~\bibnamefont{Stojanovic}},
  \bibinfo{author}{\bibfnamefont{M.}~\bibnamefont{Bogdanovic}},
  \bibinfo{author}{\bibfnamefont{S.}~\bibnamefont{Koicki}},
  \bibinfo{author}{\bibfnamefont{W.}~\bibnamefont{Gelletly}},
  \bibnamefont{et~al.}, \bibinfo{journal}{Journal of Physics G: Nuclear
  Physics} \textbf{\bibinfo{volume}{7}}, \bibinfo{pages}{455}
  (\bibinfo{year}{1981}),
  \urlprefix\url{http://stacks.iop.org/0305-4616/7/i=4/a=011}.

\bibitem[{\citenamefont{Aprahamian et~al.}(2006)\citenamefont{Aprahamian, Wu,
  Lesher, Warner, Gelletly, B{\"o}rner, Hoyler, Schreckenbach, Casten, Shi
  et~al.}}]{aprahamian2006}
\bibinfo{author}{\bibfnamefont{A.}~\bibnamefont{Aprahamian}},
  \bibinfo{author}{\bibfnamefont{X.}~\bibnamefont{Wu}},
  \bibinfo{author}{\bibfnamefont{S.~R.} \bibnamefont{Lesher}},
  \bibinfo{author}{\bibfnamefont{D.~D.} \bibnamefont{Warner}},
  \bibinfo{author}{\bibfnamefont{W.}~\bibnamefont{Gelletly}},
  \bibinfo{author}{\bibfnamefont{H.~G.} \bibnamefont{B{\"o}rner}},
  \bibinfo{author}{\bibfnamefont{F.}~\bibnamefont{Hoyler}},
  \bibinfo{author}{\bibfnamefont{K.}~\bibnamefont{Schreckenbach}},
  \bibinfo{author}{\bibfnamefont{R.~F.} \bibnamefont{Casten}},
  \bibinfo{author}{\bibfnamefont{Z.~R.} \bibnamefont{Shi}},
  \bibnamefont{et~al.}, \bibinfo{journal}{Nuclear Physics A}
  \textbf{\bibinfo{volume}{764}}, \bibinfo{pages}{42 } (\bibinfo{year}{2006}),
    \urlprefix\url{http://www.sciencedirect.com/science/article/pii/S03759474050%
1136X}.

\bibitem[{\citenamefont{Bohr}(1952)}]{bohr_1952}
\bibinfo{author}{\bibfnamefont{A.}~\bibnamefont{Bohr}}, \bibinfo{journal}{Dan.
  Mat. Fys. Medd.} \textbf{\bibinfo{volume}{26}}, \bibinfo{pages}{no. 14}
  (\bibinfo{year}{1952}).

\bibitem[{\citenamefont{Bohr and Mottelson}(1953)}]{bohrmottelson_1953}
\bibinfo{author}{\bibfnamefont{A.}~\bibnamefont{Bohr}} \bibnamefont{and}
  \bibinfo{author}{\bibfnamefont{B.~R.} \bibnamefont{Mottelson}},
  \bibinfo{journal}{Dan. Mat. Fys. Medd.} \textbf{\bibinfo{volume}{27}},
  \bibinfo{pages}{no. 16} (\bibinfo{year}{1953}).

\bibitem[{\citenamefont{Bohr and Mottelson}(1975)}]{bmbook}
\bibinfo{author}{\bibfnamefont{A.}~\bibnamefont{Bohr}} \bibnamefont{and}
  \bibinfo{author}{\bibfnamefont{B.~R.} \bibnamefont{Mottelson}},
  \emph{\bibinfo{title}{Nuclear Structure}}, vol. \bibinfo{volume}{II: Nuclear
  Deformation} (\bibinfo{publisher}{W.A. Benjamin Inc.},
  \bibinfo{address}{Reading, Massachusetts, USA}, \bibinfo{year}{1975}).

\bibitem[{\citenamefont{Faessler et~al.}(1965)\citenamefont{Faessler, Greiner,
  and Sheline}}]{Faessler1965}
\bibinfo{author}{\bibfnamefont{A.}~\bibnamefont{Faessler}},
  \bibinfo{author}{\bibfnamefont{W.}~\bibnamefont{Greiner}}, \bibnamefont{and}
  \bibinfo{author}{\bibfnamefont{R.~K.} \bibnamefont{Sheline}},
  \bibinfo{journal}{Nuclear Physics} \textbf{\bibinfo{volume}{70}},
  \bibinfo{pages}{33 } (\bibinfo{year}{1965}),   \urlprefix\url{http://www.sciencedirect.com/science/article/pii/002955826590%
2245}.

\bibitem[{\citenamefont{Eisenberg and Greiner}(1970)}]{eisenberg}
\bibinfo{author}{\bibfnamefont{J.~M.} \bibnamefont{Eisenberg}}
  \bibnamefont{and} \bibinfo{author}{\bibfnamefont{W.}~\bibnamefont{Greiner}},
  \emph{\bibinfo{title}{Nuclear Models: Collective and Single-Particle
  Phenomena}} (\bibinfo{publisher}{North-Holland Publishing Company Ltd.},
  \bibinfo{address}{London}, \bibinfo{year}{1970}).

\bibitem[{\citenamefont{Gneuss and Greiner}(1971)}]{Gneuss1971}
\bibinfo{author}{\bibfnamefont{G.}~\bibnamefont{Gneuss}} \bibnamefont{and}
  \bibinfo{author}{\bibfnamefont{W.}~\bibnamefont{Greiner}},
  \bibinfo{journal}{Nuclear Physics A} \textbf{\bibinfo{volume}{171}},
  \bibinfo{pages}{449 } (\bibinfo{year}{1971}), 
  \urlprefix\url{http://www.sciencedirect.com/science/article/pii/037594747190%
5963}.

\bibitem[{\citenamefont{Hess et~al.}(1980)\citenamefont{Hess, Seiwert, Maruhn,
  and Greiner}}]{hess1980}
\bibinfo{author}{\bibfnamefont{P.}~\bibnamefont{Hess}},
  \bibinfo{author}{\bibfnamefont{M.}~\bibnamefont{Seiwert}},
  \bibinfo{author}{\bibfnamefont{J.}~\bibnamefont{Maruhn}}, \bibnamefont{and}
  \bibinfo{author}{\bibfnamefont{W.}~\bibnamefont{Greiner}},
  \bibinfo{journal}{Zeitschrift f{\"u}r Physik A Atoms and Nuclei}
  \textbf{\bibinfo{volume}{296}}, \bibinfo{pages}{147} (\bibinfo{year}{1980}),
  \urlprefix\url{http://dx.doi.org/10.1007/BF01412656}.

\bibitem[{\citenamefont{Arima and Iachello}(1975)}]{arima1975}
\bibinfo{author}{\bibfnamefont{A.}~\bibnamefont{Arima}} \bibnamefont{and}
  \bibinfo{author}{\bibfnamefont{F.}~\bibnamefont{Iachello}},
  \bibinfo{journal}{Phys. Rev. Lett.} \textbf{\bibinfo{volume}{35}},
  \bibinfo{pages}{1069} (\bibinfo{year}{1975}),
  \urlprefix\url{http://link.aps.org/doi/10.1103/PhysRevLett.35.1069}.

\bibitem[{\citenamefont{Iachello and Arima}(1987)}]{iachello}
\bibinfo{author}{\bibfnamefont{F.}~\bibnamefont{Iachello}} \bibnamefont{and}
  \bibinfo{author}{\bibfnamefont{A.}~\bibnamefont{Arima}},
  \emph{\bibinfo{title}{The Interacting Boson Model}}
  (\bibinfo{publisher}{Cambridge University Press},
  \bibinfo{address}{Cambridge, UK}, \bibinfo{year}{1987}).

\bibitem[{\citenamefont{Kerman}(1956)}]{kerman1956}
\bibinfo{author}{\bibfnamefont{A.~K.} \bibnamefont{Kerman}},
  \bibinfo{journal}{Dan. Mat. Fys. Medd.} \textbf{\bibinfo{volume}{30}},
  \bibinfo{pages}{no. 15} (\bibinfo{year}{1956}).

\bibitem[{\citenamefont{Iachello and Scholten}(1979)}]{iachello1979}
\bibinfo{author}{\bibfnamefont{F.}~\bibnamefont{Iachello}} \bibnamefont{and}
  \bibinfo{author}{\bibfnamefont{O.}~\bibnamefont{Scholten}},
  \bibinfo{journal}{Phys. Rev. Lett.} \textbf{\bibinfo{volume}{43}},
  \bibinfo{pages}{679} (\bibinfo{year}{1979}),
  \urlprefix\url{http://link.aps.org/doi/10.1103/PhysRevLett.43.679}.

\bibitem[{\citenamefont{{\AA}berg et~al.}(1990)\citenamefont{{\AA}berg,
  Flocard, and Nazarewicz}}]{aberg1990}
\bibinfo{author}{\bibfnamefont{S.}~\bibnamefont{{\AA}berg}},
  \bibinfo{author}{\bibfnamefont{H.}~\bibnamefont{Flocard}}, \bibnamefont{and}
  \bibinfo{author}{\bibfnamefont{W.}~\bibnamefont{Nazarewicz}},
  \bibinfo{journal}{Annual Review of Nuclear and Particle Science}
  \textbf{\bibinfo{volume}{40}}, \bibinfo{pages}{439} (\bibinfo{year}{1990}),
  \urlprefix\url{http://www.annualreviews.org/doi/abs/10.1146/annurev.ns.40.12%
0190.002255}.

\bibitem[{\citenamefont{Frauendorf}(2001)}]{frauendorf2001}
\bibinfo{author}{\bibfnamefont{S.}~\bibnamefont{Frauendorf}},
  \bibinfo{journal}{Rev. Mod. Phys.} \textbf{\bibinfo{volume}{73}},
  \bibinfo{pages}{463} (\bibinfo{year}{2001}),
  \urlprefix\url{http://link.aps.org/doi/10.1103/RevModPhys.73.463}.

\bibitem[{\citenamefont{Vargas et~al.}(2000)\citenamefont{Vargas, Hirsch,
  Beuschel, and Draayer}}]{vargas2000}
\bibinfo{author}{\bibfnamefont{C.}~\bibnamefont{Vargas}},
  \bibinfo{author}{\bibfnamefont{J.~G.} \bibnamefont{Hirsch}},
  \bibinfo{author}{\bibfnamefont{T.}~\bibnamefont{Beuschel}}, \bibnamefont{and}
  \bibinfo{author}{\bibfnamefont{J.~P.} \bibnamefont{Draayer}},
  \bibinfo{journal}{Phys. Rev. C} \textbf{\bibinfo{volume}{61}},
  \bibinfo{pages}{031301} (\bibinfo{year}{2000}),
  \urlprefix\url{http://link.aps.org/doi/10.1103/PhysRevC.61.031301}.

\bibitem[{\citenamefont{Halbert and Nazarewicz}(1993)}]{halbert1993}
\bibinfo{author}{\bibfnamefont{E.~C.} \bibnamefont{Halbert}} \bibnamefont{and}
  \bibinfo{author}{\bibfnamefont{W.}~\bibnamefont{Nazarewicz}},
  \bibinfo{journal}{Phys. Rev. C} \textbf{\bibinfo{volume}{48}},
  \bibinfo{pages}{R2158} (\bibinfo{year}{1993}),
  \urlprefix\url{http://link.aps.org/doi/10.1103/PhysRevC.48.R2158}.

\bibitem[{\citenamefont{Dracoulis et~al.}(1998)\citenamefont{Dracoulis, Kondev,
  and Walker}}]{dracoulis1998}
\bibinfo{author}{\bibfnamefont{G.}~\bibnamefont{Dracoulis}},
  \bibinfo{author}{\bibfnamefont{F.}~\bibnamefont{Kondev}}, \bibnamefont{and}
  \bibinfo{author}{\bibfnamefont{P.}~\bibnamefont{Walker}},
  \bibinfo{journal}{Physics Letters B} \textbf{\bibinfo{volume}{419}},
  \bibinfo{pages}{7 } (\bibinfo{year}{1998}), 
  \urlprefix\url{http://www.sciencedirect.com/science/article/pii/S03702693970%
14561}.

\bibitem[{\citenamefont{Zhang et~al.}(2009)\citenamefont{Zhang, Wu, Lei, and
  Zeng}}]{zhang2009}
\bibinfo{author}{\bibfnamefont{Z.}~\bibnamefont{Zhang}},
  \bibinfo{author}{\bibfnamefont{X.}~\bibnamefont{Wu}},
  \bibinfo{author}{\bibfnamefont{Y.}~\bibnamefont{Lei}}, \bibnamefont{and}
  \bibinfo{author}{\bibfnamefont{J.}~\bibnamefont{Zeng}},
  \bibinfo{journal}{Nuclear Physics A} \textbf{\bibinfo{volume}{816}},
  \bibinfo{pages}{19 } (\bibinfo{year}{2009}), 
  \urlprefix\url{http://www.sciencedirect.com/science/article/pii/S03759474080%
07550}.

\bibitem[{\citenamefont{Wu et~al.}(2011)\citenamefont{Wu, Zhang, Zeng, and
  Lei}}]{wu2011}
\bibinfo{author}{\bibfnamefont{X.}~\bibnamefont{Wu}},
  \bibinfo{author}{\bibfnamefont{Z.~H.} \bibnamefont{Zhang}},
  \bibinfo{author}{\bibfnamefont{J.~Y.} \bibnamefont{Zeng}}, \bibnamefont{and}
  \bibinfo{author}{\bibfnamefont{Y.~A.} \bibnamefont{Lei}},
  \bibinfo{journal}{Phys. Rev. C} \textbf{\bibinfo{volume}{83}},
  \bibinfo{pages}{034323} (\bibinfo{year}{2011}),
  \urlprefix\url{http://link.aps.org/doi/10.1103/PhysRevC.83.034323}.

\bibitem[{\citenamefont{Rowe}(2004)}]{Rowe2004}
\bibinfo{author}{\bibfnamefont{D.~J.} \bibnamefont{Rowe}},
  \bibinfo{journal}{Nuclear Physics A} \textbf{\bibinfo{volume}{735}},
  \bibinfo{pages}{372 } (\bibinfo{year}{2004}), 
  \urlprefix\url{http://www.sciencedirect.com/science/article/pii/S03759474040%
02076}.

\bibitem[{\citenamefont{Rowe et~al.}(2009)\citenamefont{Rowe, Welsh, and
  Caprio}}]{rowe2009}
\bibinfo{author}{\bibfnamefont{D.~J.} \bibnamefont{Rowe}},
  \bibinfo{author}{\bibfnamefont{T.~A.} \bibnamefont{Welsh}}, \bibnamefont{and}
  \bibinfo{author}{\bibfnamefont{M.~A.} \bibnamefont{Caprio}},
  \bibinfo{journal}{Phys. Rev. C} \textbf{\bibinfo{volume}{79}},
  \bibinfo{pages}{054304} (\bibinfo{year}{2009}),
  \urlprefix\url{http://link.aps.org/doi/10.1103/PhysRevC.79.054304}.

\bibitem[{\citenamefont{Matsuyanagi et~al.}(2010)\citenamefont{Matsuyanagi,
  Matsuo, Nakatsukasa, Hinohara, and Sato}}]{matsuyanagi2010}
\bibinfo{author}{\bibfnamefont{K.}~\bibnamefont{Matsuyanagi}},
  \bibinfo{author}{\bibfnamefont{M.}~\bibnamefont{Matsuo}},
  \bibinfo{author}{\bibfnamefont{T.}~\bibnamefont{Nakatsukasa}},
  \bibinfo{author}{\bibfnamefont{N.}~\bibnamefont{Hinohara}}, \bibnamefont{and}
  \bibinfo{author}{\bibfnamefont{K.}~\bibnamefont{Sato}},
  \bibinfo{journal}{Journal of Physics G: Nuclear and Particle Physics}
  \textbf{\bibinfo{volume}{37}}, \bibinfo{pages}{064018}
  (\bibinfo{year}{2010}),
  \urlprefix\url{http://stacks.iop.org/0954-3899/37/i=6/a=064018}.

\bibitem[{\citenamefont{Papenbrock}(2011)}]{papenbrock2011}
\bibinfo{author}{\bibfnamefont{T.}~\bibnamefont{Papenbrock}},
  \bibinfo{journal}{Nuclear Physics A} \textbf{\bibinfo{volume}{852}},
  \bibinfo{pages}{36 } (\bibinfo{year}{2011}), 
  \urlprefix\url{http://www.sciencedirect.com/science/article/pii/S03759474100%
07773}.

\bibitem[{\citenamefont{Weinberg}(1990)}]{Weinberg1990}
\bibinfo{author}{\bibfnamefont{S.}~\bibnamefont{Weinberg}},
  \bibinfo{journal}{Physics Letters B} \textbf{\bibinfo{volume}{251}},
  \bibinfo{pages}{288 } (\bibinfo{year}{1990}), 
  \urlprefix\url{http://www.sciencedirect.com/science/article/pii/037026939090%
9383}.

\bibitem[{\citenamefont{Weinberg}(1991)}]{Weinberg1991}
\bibinfo{author}{\bibfnamefont{S.}~\bibnamefont{Weinberg}},
  \bibinfo{journal}{Nuclear Physics B} \textbf{\bibinfo{volume}{363}},
  \bibinfo{pages}{3 } (\bibinfo{year}{1991}), 
  \urlprefix\url{http://www.sciencedirect.com/science/article/pii/055032139190%
231L}.

\bibitem[{\citenamefont{van Kolck}(1994)}]{vankolck1994}
\bibinfo{author}{\bibfnamefont{U.}~\bibnamefont{van Kolck}},
  \bibinfo{journal}{Phys. Rev. C} \textbf{\bibinfo{volume}{49}},
  \bibinfo{pages}{2932} (\bibinfo{year}{1994}),
  \urlprefix\url{http://link.aps.org/doi/10.1103/PhysRevC.49.2932}.

\bibitem[{\citenamefont{{Bedaque} and {van Kolck}}(2002)}]{bedaque2002}
\bibinfo{author}{\bibfnamefont{P.~F.} \bibnamefont{{Bedaque}}}
  \bibnamefont{and} \bibinfo{author}{\bibfnamefont{U.}~\bibnamefont{{van
  Kolck}}}, \bibinfo{journal}{Annual Review of Nuclear and Particle Science}
  \textbf{\bibinfo{volume}{52}}, \bibinfo{pages}{339} (\bibinfo{year}{2002}),
  \eprint{arXiv:nucl-th/0203055}.

\bibitem[{\citenamefont{Epelbaum}(2006)}]{Epelbaum2006}
\bibinfo{author}{\bibfnamefont{E.}~\bibnamefont{Epelbaum}},
  \bibinfo{journal}{Progress in Particle and Nuclear Physics}
  \textbf{\bibinfo{volume}{57}}, \bibinfo{pages}{654 } (\bibinfo{year}{2006}),
    \urlprefix\url{http://www.sciencedirect.com/science/article/pii/S01466410050%
01018}.

\bibitem[{\citenamefont{Machleidt and Entem}(2011)}]{Machleidt2011}
\bibinfo{author}{\bibfnamefont{R.}~\bibnamefont{Machleidt}} \bibnamefont{and}
  \bibinfo{author}{\bibfnamefont{D.}~\bibnamefont{Entem}},
  \bibinfo{journal}{Physics Reports} \textbf{\bibinfo{volume}{503}},
  \bibinfo{pages}{1 } (\bibinfo{year}{2011}), 
  \urlprefix\url{http://www.sciencedirect.com/science/article/pii/S03701573110%
00457}.

\bibitem[{\citenamefont{Bertulani et~al.}(2002)\citenamefont{Bertulani, Hammer,
  and van Kolck}}]{bertulani2002}
\bibinfo{author}{\bibfnamefont{C.}~\bibnamefont{Bertulani}},
  \bibinfo{author}{\bibfnamefont{H.-W.} \bibnamefont{Hammer}},
  \bibnamefont{and} \bibinfo{author}{\bibfnamefont{U.}~\bibnamefont{van
  Kolck}}, \bibinfo{journal}{Nuclear Physics A} \textbf{\bibinfo{volume}{712}},
  \bibinfo{pages}{37 } (\bibinfo{year}{2002}), 
  \urlprefix\url{http://www.sciencedirect.com/science/article/pii/S03759474020%
12708}.

\bibitem[{\citenamefont{Higa et~al.}(2008)\citenamefont{Higa, Hammer, and van
  Kolck}}]{higa2008}
\bibinfo{author}{\bibfnamefont{R.}~\bibnamefont{Higa}},
  \bibinfo{author}{\bibfnamefont{H.-W.} \bibnamefont{Hammer}},
  \bibnamefont{and} \bibinfo{author}{\bibfnamefont{U.}~\bibnamefont{van
  Kolck}}, \bibinfo{journal}{Nuclear Physics A} \textbf{\bibinfo{volume}{809}},
  \bibinfo{pages}{171 } (\bibinfo{year}{2008}), 
  \urlprefix\url{http://www.sciencedirect.com/science/article/pii/S03759474080%
05757}.

\bibitem[{\citenamefont{Marini et~al.}(1998)\citenamefont{Marini, Pistolesi,
  and Strinati}}]{marini1998}
\bibinfo{author}{\bibfnamefont{M.}~\bibnamefont{Marini}},
  \bibinfo{author}{\bibfnamefont{F.}~\bibnamefont{Pistolesi}},
  \bibnamefont{and} \bibinfo{author}{\bibfnamefont{G.}~\bibnamefont{Strinati}},
  \bibinfo{journal}{The European Physical Journal B - Condensed Matter and
  Complex Systems} \textbf{\bibinfo{volume}{1}}, \bibinfo{pages}{151}
  (\bibinfo{year}{1998}), 
  \urlprefix\url{http://dx.doi.org/10.1007/s100510050165}.

\bibitem[{\citenamefont{Papenbrock and Bertsch}(1999)}]{papenbrock1999}
\bibinfo{author}{\bibfnamefont{T.}~\bibnamefont{Papenbrock}} \bibnamefont{and}
  \bibinfo{author}{\bibfnamefont{G.~F.} \bibnamefont{Bertsch}},
  \bibinfo{journal}{Phys. Rev. C} \textbf{\bibinfo{volume}{59}},
  \bibinfo{pages}{2052} (\bibinfo{year}{1999}),
  \urlprefix\url{http://link.aps.org/doi/10.1103/PhysRevC.59.2052}.

\bibitem[{\citenamefont{Hammer and Furnstahl}(2000)}]{hammer2000}
\bibinfo{author}{\bibfnamefont{H.-W.} \bibnamefont{Hammer}} \bibnamefont{and}
  \bibinfo{author}{\bibfnamefont{R.}~\bibnamefont{Furnstahl}},
  \bibinfo{journal}{Nuclear Physics A} \textbf{\bibinfo{volume}{678}},
  \bibinfo{pages}{277 } (\bibinfo{year}{2000}), 
  \urlprefix\url{http://www.sciencedirect.com/science/article/pii/S03759474000%
03250}.

\bibitem[{\citenamefont{Furnstahl et~al.}(2007)\citenamefont{Furnstahl, Hammer,
  and Puglia}}]{Furnstahl2007}
\bibinfo{author}{\bibfnamefont{R.}~\bibnamefont{Furnstahl}},
  \bibinfo{author}{\bibfnamefont{H.-W.} \bibnamefont{Hammer}},
  \bibnamefont{and} \bibinfo{author}{\bibfnamefont{S.}~\bibnamefont{Puglia}},
  \bibinfo{journal}{Annals of Physics} \textbf{\bibinfo{volume}{322}},
  \bibinfo{pages}{2703 } (\bibinfo{year}{2007}), 
  \urlprefix\url{http://www.sciencedirect.com/science/article/pii/S00034916070%
00085}.

\bibitem[{\citenamefont{Warner et~al.}(1981)\citenamefont{Warner, Casten, and
  Davidson}}]{warner1981}
\bibinfo{author}{\bibfnamefont{D.~D.} \bibnamefont{Warner}},
  \bibinfo{author}{\bibfnamefont{R.~F.} \bibnamefont{Casten}},
  \bibnamefont{and} \bibinfo{author}{\bibfnamefont{W.~F.}
  \bibnamefont{Davidson}}, \bibinfo{journal}{Phys. Rev. C}
  \textbf{\bibinfo{volume}{24}}, \bibinfo{pages}{1713} (\bibinfo{year}{1981}),
  \urlprefix\url{http://link.aps.org/doi/10.1103/PhysRevC.24.1713}.

\bibitem[{\citenamefont{Grosse et~al.}(1981)\citenamefont{Grosse, Balanda,
  Emling, Folkmann, Fuchs, Piercey, Schwalm, Simon, Wollersheim, Evers
  et~al.}}]{Grosse1981}
\bibinfo{author}{\bibfnamefont{E.}~\bibnamefont{Grosse}},
  \bibinfo{author}{\bibfnamefont{A.}~\bibnamefont{Balanda}},
  \bibinfo{author}{\bibfnamefont{H.}~\bibnamefont{Emling}},
  \bibinfo{author}{\bibfnamefont{F.}~\bibnamefont{Folkmann}},
  \bibinfo{author}{\bibfnamefont{P.}~\bibnamefont{Fuchs}},
  \bibinfo{author}{\bibfnamefont{R.~B.} \bibnamefont{Piercey}},
  \bibinfo{author}{\bibfnamefont{D.}~\bibnamefont{Schwalm}},
  \bibinfo{author}{\bibfnamefont{R.~S.} \bibnamefont{Simon}},
  \bibinfo{author}{\bibfnamefont{H.~J.} \bibnamefont{Wollersheim}},
  \bibinfo{author}{\bibfnamefont{D.}~\bibnamefont{Evers}},
  \bibnamefont{et~al.}, \bibinfo{journal}{Physica Scripta}
  \textbf{\bibinfo{volume}{24}}, \bibinfo{pages}{337} (\bibinfo{year}{1981}),
  \urlprefix\url{http://stacks.iop.org/1402-4896/24/i=1B/a=033}.

\bibitem[{\citenamefont{Kuyucak and Morrison}(1988)}]{kuyucak1988}
\bibinfo{author}{\bibfnamefont{S.}~\bibnamefont{Kuyucak}} \bibnamefont{and}
  \bibinfo{author}{\bibfnamefont{I.}~\bibnamefont{Morrison}},
  \bibinfo{journal}{Phys. Rev. C} \textbf{\bibinfo{volume}{38}},
  \bibinfo{pages}{2482} (\bibinfo{year}{1988}),
  \urlprefix\url{http://link.aps.org/doi/10.1103/PhysRevC.38.2482}.

\bibitem[{\citenamefont{Caprio}(2005)}]{caprio2005}
\bibinfo{author}{\bibfnamefont{M.~A.} \bibnamefont{Caprio}},
  \bibinfo{journal}{Phys. Rev. C} \textbf{\bibinfo{volume}{72}},
  \bibinfo{pages}{054323} (\bibinfo{year}{2005}),
  \urlprefix\url{http://link.aps.org/doi/10.1103/PhysRevC.72.054323}.

\bibitem[{\citenamefont{Caprio}(2009)}]{Caprio2009}
\bibinfo{author}{\bibfnamefont{M.}~\bibnamefont{Caprio}},
  \bibinfo{journal}{Physics Letters B} \textbf{\bibinfo{volume}{672}},
  \bibinfo{pages}{396 } (\bibinfo{year}{2009}), 
  \urlprefix\url{http://www.sciencedirect.com/science/article/pii/S03702693090%
01038}.

\bibitem[{\citenamefont{Caprio}(2011)}]{Caprio2011}
\bibinfo{author}{\bibfnamefont{M.~A.} \bibnamefont{Caprio}},
  \bibinfo{journal}{Phys. Rev. C} \textbf{\bibinfo{volume}{83}},
  \bibinfo{pages}{064309} (\bibinfo{year}{2011}),
  \urlprefix\url{http://link.aps.org/doi/10.1103/PhysRevC.83.064309}.

\bibitem[{\citenamefont{Heyde and Wood}(2011)}]{heyde2011}
\bibinfo{author}{\bibfnamefont{K.}~\bibnamefont{Heyde}} \bibnamefont{and}
  \bibinfo{author}{\bibfnamefont{J.~L.} \bibnamefont{Wood}},
  \bibinfo{journal}{Rev. Mod. Phys.} \textbf{\bibinfo{volume}{83}},
  \bibinfo{pages}{1467} (\bibinfo{year}{2011}),
  \urlprefix\url{http://link.aps.org/doi/10.1103/RevModPhys.83.1467}.

\bibitem[{\citenamefont{Iachello}(2001)}]{iachello2001}
\bibinfo{author}{\bibfnamefont{F.}~\bibnamefont{Iachello}},
  \bibinfo{journal}{Phys. Rev. Lett.} \textbf{\bibinfo{volume}{87}},
  \bibinfo{pages}{052502} (\bibinfo{year}{2001}),
  \urlprefix\url{http://link.aps.org/doi/10.1103/PhysRevLett.87.052502}.

\bibitem[{\citenamefont{Casten and Zamfir}(2001)}]{casten2001}
\bibinfo{author}{\bibfnamefont{R.~F.} \bibnamefont{Casten}} \bibnamefont{and}
  \bibinfo{author}{\bibfnamefont{N.~V.} \bibnamefont{Zamfir}},
  \bibinfo{journal}{Phys. Rev. Lett.} \textbf{\bibinfo{volume}{87}},
  \bibinfo{pages}{052503} (\bibinfo{year}{2001}),
  \urlprefix\url{http://link.aps.org/doi/10.1103/PhysRevLett.87.052503}.

\bibitem[{\citenamefont{Pietralla and Gorbachenko}(2004)}]{pietralla2004}
\bibinfo{author}{\bibfnamefont{N.}~\bibnamefont{Pietralla}} \bibnamefont{and}
  \bibinfo{author}{\bibfnamefont{O.~M.} \bibnamefont{Gorbachenko}},
  \bibinfo{journal}{Phys. Rev. C} \textbf{\bibinfo{volume}{70}},
  \bibinfo{pages}{011304} (\bibinfo{year}{2004}),
  \urlprefix\url{http://link.aps.org/doi/10.1103/PhysRevC.70.011304}.

\bibitem[{\citenamefont{Weinberg}(1996)}]{weinbergbook}
\bibinfo{author}{\bibfnamefont{S.}~\bibnamefont{Weinberg}},
  \emph{\bibinfo{title}{The Quantum Theory of Fields, Vol.II}}
  (\bibinfo{publisher}{Cambridge University Press},
  \bibinfo{address}{Cambridge, UK}, \bibinfo{year}{1996}).

\bibitem[{\citenamefont{Bohr and Mottelson}(1982)}]{bohrmottelson1982}
\bibinfo{author}{\bibfnamefont{A.}~\bibnamefont{Bohr}} \bibnamefont{and}
  \bibinfo{author}{\bibfnamefont{B.~R.} \bibnamefont{Mottelson}},
  \bibinfo{journal}{Physica Scripta} \textbf{\bibinfo{volume}{25}},
  \bibinfo{pages}{28} (\bibinfo{year}{1982}),
  \urlprefix\url{http://stacks.iop.org/1402-4896/25/i=1A/a=005}.

\bibitem[{\citenamefont{Fukuda}(1988)}]{fukuda1988}
\bibinfo{author}{\bibfnamefont{R.}~\bibnamefont{Fukuda}},
  \bibinfo{journal}{Phys. Rev. Lett.} \textbf{\bibinfo{volume}{61}},
  \bibinfo{pages}{1549} (\bibinfo{year}{1988}),
  \urlprefix\url{http://link.aps.org/doi/10.1103/PhysRevLett.61.1549}.

\bibitem[{\citenamefont{Inagaki and Fukuda}(1992)}]{Inagaki1992}
\bibinfo{author}{\bibfnamefont{T.}~\bibnamefont{Inagaki}} \bibnamefont{and}
  \bibinfo{author}{\bibfnamefont{R.}~\bibnamefont{Fukuda}},
  \bibinfo{journal}{Phys. Rev. B} \textbf{\bibinfo{volume}{46}},
  \bibinfo{pages}{10931} (\bibinfo{year}{1992}),
  \urlprefix\url{http://link.aps.org/doi/10.1103/PhysRevB.46.10931}.

\bibitem[{\citenamefont{Fukuda et~al.}(1994)\citenamefont{Fukuda, Kotani,
  Suzuki, and Yokojima}}]{fukuda1994}
\bibinfo{author}{\bibfnamefont{R.}~\bibnamefont{Fukuda}},
  \bibinfo{author}{\bibfnamefont{T.}~\bibnamefont{Kotani}},
  \bibinfo{author}{\bibfnamefont{Y.}~\bibnamefont{Suzuki}}, \bibnamefont{and}
  \bibinfo{author}{\bibfnamefont{S.}~\bibnamefont{Yokojima}},
  \bibinfo{journal}{Progress of Theoretical Physics}
  \textbf{\bibinfo{volume}{92}}, \bibinfo{pages}{833} (\bibinfo{year}{1994}),
  \urlprefix\url{http://ptp.ipap.jp/link?PTP/92/833/}.

\bibitem[{\citenamefont{Sood et~al.}(1991)\citenamefont{Sood, Headly, and
  Sheline}}]{Sood1991}
\bibinfo{author}{\bibfnamefont{P.}~\bibnamefont{Sood}},
  \bibinfo{author}{\bibfnamefont{D.}~\bibnamefont{Headly}}, \bibnamefont{and}
  \bibinfo{author}{\bibfnamefont{R.}~\bibnamefont{Sheline}},
  \bibinfo{journal}{Atomic Data and Nuclear Data Tables}
  \textbf{\bibinfo{volume}{47}}, \bibinfo{pages}{89 } (\bibinfo{year}{1991}),
    \urlprefix\url{http://www.sciencedirect.com/science/article/pii/0092640X9190%
019Z}.

\bibitem[{\citenamefont{Sood et~al.}(1992)\citenamefont{Sood, Headly, and
  Sheline}}]{Sood1992}
\bibinfo{author}{\bibfnamefont{P.}~\bibnamefont{Sood}},
  \bibinfo{author}{\bibfnamefont{D.}~\bibnamefont{Headly}}, \bibnamefont{and}
  \bibinfo{author}{\bibfnamefont{R.}~\bibnamefont{Sheline}},
  \bibinfo{journal}{Atomic Data and Nuclear Data Tables}
  \textbf{\bibinfo{volume}{51}}, \bibinfo{pages}{273 } (\bibinfo{year}{1992}),
    \urlprefix\url{http://www.sciencedirect.com/science/article/pii/0092640X9290%
003Z}.

\bibitem[{\citenamefont{B\"orner et~al.}(1991)\citenamefont{B\"orner, Jolie,
  Robinson, Krusche, Piepenbring, Casten, Aprahamian, and
  Draayer}}]{Boerner1991}
\bibinfo{author}{\bibfnamefont{H.~G.} \bibnamefont{B\"orner}},
  \bibinfo{author}{\bibfnamefont{J.}~\bibnamefont{Jolie}},
  \bibinfo{author}{\bibfnamefont{S.~J.} \bibnamefont{Robinson}},
  \bibinfo{author}{\bibfnamefont{B.}~\bibnamefont{Krusche}},
  \bibinfo{author}{\bibfnamefont{R.}~\bibnamefont{Piepenbring}},
  \bibinfo{author}{\bibfnamefont{R.~F.} \bibnamefont{Casten}},
  \bibinfo{author}{\bibfnamefont{A.}~\bibnamefont{Aprahamian}},
  \bibnamefont{and} \bibinfo{author}{\bibfnamefont{J.~P.}
  \bibnamefont{Draayer}}, \bibinfo{journal}{Phys. Rev. Lett.}
  \textbf{\bibinfo{volume}{66}}, \bibinfo{pages}{691} (\bibinfo{year}{1991}),
  \urlprefix\url{http://link.aps.org/doi/10.1103/PhysRevLett.66.691}.

\bibitem[{\citenamefont{Oshima et~al.}(1995)\citenamefont{Oshima, Morikawa,
  Hatsukawa, Ichikawa, Shinohara, Matsuo, Kusakari, Kobayashi, Sugawara, and
  Inamura}}]{Oshima1995}
\bibinfo{author}{\bibfnamefont{M.}~\bibnamefont{Oshima}},
  \bibinfo{author}{\bibfnamefont{T.}~\bibnamefont{Morikawa}},
  \bibinfo{author}{\bibfnamefont{Y.}~\bibnamefont{Hatsukawa}},
  \bibinfo{author}{\bibfnamefont{S.}~\bibnamefont{Ichikawa}},
  \bibinfo{author}{\bibfnamefont{N.}~\bibnamefont{Shinohara}},
  \bibinfo{author}{\bibfnamefont{M.}~\bibnamefont{Matsuo}},
  \bibinfo{author}{\bibfnamefont{H.}~\bibnamefont{Kusakari}},
  \bibinfo{author}{\bibfnamefont{N.}~\bibnamefont{Kobayashi}},
  \bibinfo{author}{\bibfnamefont{M.}~\bibnamefont{Sugawara}}, \bibnamefont{and}
  \bibinfo{author}{\bibfnamefont{T.}~\bibnamefont{Inamura}},
  \bibinfo{journal}{Phys. Rev. C} \textbf{\bibinfo{volume}{52}},
  \bibinfo{pages}{3492} (\bibinfo{year}{1995}),
  \urlprefix\url{http://link.aps.org/doi/10.1103/PhysRevC.52.3492}.

\bibitem[{\citenamefont{H{\"a}rtlein et~al.}(1998)\citenamefont{H{\"a}rtlein,
  Heinebrodt, Schwalm, and Fahlander}}]{Haertlein1998}
\bibinfo{author}{\bibfnamefont{T.}~\bibnamefont{H{\"a}rtlein}},
  \bibinfo{author}{\bibfnamefont{M.}~\bibnamefont{Heinebrodt}},
  \bibinfo{author}{\bibfnamefont{D.}~\bibnamefont{Schwalm}}, \bibnamefont{and}
  \bibinfo{author}{\bibfnamefont{C.}~\bibnamefont{Fahlander}},
  \bibinfo{journal}{The European Physical Journal A - Hadrons and Nuclei}
  \textbf{\bibinfo{volume}{2}}, \bibinfo{pages}{253} (\bibinfo{year}{1998}),
    \urlprefix\url{http://dx.doi.org/10.1007/s100500050117}.

\bibitem[{\citenamefont{Fahlander et~al.}(1996)\citenamefont{Fahlander,
  Axelsson, Heinebrodt, H{\"a}rtlein, and Schwalm}}]{Fahlander1996}
\bibinfo{author}{\bibfnamefont{C.}~\bibnamefont{Fahlander}},
  \bibinfo{author}{\bibfnamefont{A.}~\bibnamefont{Axelsson}},
  \bibinfo{author}{\bibfnamefont{M.}~\bibnamefont{Heinebrodt}},
  \bibinfo{author}{\bibfnamefont{T.}~\bibnamefont{H{\"a}rtlein}},
  \bibnamefont{and} \bibinfo{author}{\bibfnamefont{D.}~\bibnamefont{Schwalm}},
  \bibinfo{journal}{Physics Letters B} \textbf{\bibinfo{volume}{388}},
  \bibinfo{pages}{475 } (\bibinfo{year}{1996}), 
  \urlprefix\url{http://www.sciencedirect.com/science/article/pii/S03702693960%
12038}.

\bibitem[{\citenamefont{Garrett et~al.}(1997)\citenamefont{Garrett, Kadi, Li,
  McGrath, Sorokin, Yeh, and Yates}}]{Garrett1997}
\bibinfo{author}{\bibfnamefont{P.~E.} \bibnamefont{Garrett}},
  \bibinfo{author}{\bibfnamefont{M.}~\bibnamefont{Kadi}},
  \bibinfo{author}{\bibfnamefont{M.}~\bibnamefont{Li}},
  \bibinfo{author}{\bibfnamefont{C.~A.} \bibnamefont{McGrath}},
  \bibinfo{author}{\bibfnamefont{V.}~\bibnamefont{Sorokin}},
  \bibinfo{author}{\bibfnamefont{M.}~\bibnamefont{Yeh}}, \bibnamefont{and}
  \bibinfo{author}{\bibfnamefont{S.~W.} \bibnamefont{Yates}},
  \bibinfo{journal}{Phys. Rev. Lett.} \textbf{\bibinfo{volume}{78}},
  \bibinfo{pages}{4545} (\bibinfo{year}{1997}),
  \urlprefix\url{http://link.aps.org/doi/10.1103/PhysRevLett.78.4545}.

\bibitem[{\citenamefont{Korten et~al.}(1993)\citenamefont{Korten, H{\"a}rtlein,
  Gerl, Habs, and Schwalm}}]{Korten1993}
\bibinfo{author}{\bibfnamefont{W.}~\bibnamefont{Korten}},
  \bibinfo{author}{\bibfnamefont{T.}~\bibnamefont{H{\"a}rtlein}},
  \bibinfo{author}{\bibfnamefont{J.}~\bibnamefont{Gerl}},
  \bibinfo{author}{\bibfnamefont{D.}~\bibnamefont{Habs}}, \bibnamefont{and}
  \bibinfo{author}{\bibfnamefont{D.}~\bibnamefont{Schwalm}},
  \bibinfo{journal}{Physics Letters B} \textbf{\bibinfo{volume}{317}},
  \bibinfo{pages}{19 } (\bibinfo{year}{1993}), 
  \urlprefix\url{http://www.sciencedirect.com/science/article/pii/037026939391%
5633}.

\bibitem[{\citenamefont{Martin et~al.}(2000)\citenamefont{Martin, Garrett,
  Kadi, Warr, McEllistrem, and Yates}}]{martin2000}
\bibinfo{author}{\bibfnamefont{A.}~\bibnamefont{Martin}},
  \bibinfo{author}{\bibfnamefont{P.~E.} \bibnamefont{Garrett}},
  \bibinfo{author}{\bibfnamefont{M.}~\bibnamefont{Kadi}},
  \bibinfo{author}{\bibfnamefont{N.}~\bibnamefont{Warr}},
  \bibinfo{author}{\bibfnamefont{M.~T.} \bibnamefont{McEllistrem}},
  \bibnamefont{and} \bibinfo{author}{\bibfnamefont{S.~W.} \bibnamefont{Yates}},
  \bibinfo{journal}{Phys. Rev. C} \textbf{\bibinfo{volume}{62}},
  \bibinfo{pages}{067302} (\bibinfo{year}{2000}),
  \urlprefix\url{http://link.aps.org/doi/10.1103/PhysRevC.62.067302}.

\bibitem[{\citenamefont{Matsuo}(1984)}]{matsuo1984}
\bibinfo{author}{\bibfnamefont{M.}~\bibnamefont{Matsuo}},
  \bibinfo{journal}{Progress of Theoretical Physics}
  \textbf{\bibinfo{volume}{72}}, \bibinfo{pages}{666} (\bibinfo{year}{1984}),
  \urlprefix\url{http://ptp.ipap.jp/link?PTP/72/666/}.

\bibitem[{\citenamefont{Matsuo and Matsuyanagi}(1985)}]{matsuo1985}
\bibinfo{author}{\bibfnamefont{M.}~\bibnamefont{Matsuo}} \bibnamefont{and}
  \bibinfo{author}{\bibfnamefont{K.}~\bibnamefont{Matsuyanagi}},
  \bibinfo{journal}{Progress of Theoretical Physics}
  \textbf{\bibinfo{volume}{74}}, \bibinfo{pages}{1227} (\bibinfo{year}{1985}),
  \urlprefix\url{http://ptp.ipap.jp/link?PTP/74/1227/}.

\bibitem[{\citenamefont{Piepenbring and Jammari}(1988)}]{Piepenbring1988}
\bibinfo{author}{\bibfnamefont{R.}~\bibnamefont{Piepenbring}} \bibnamefont{and}
  \bibinfo{author}{\bibfnamefont{M.}~\bibnamefont{Jammari}},
  \bibinfo{journal}{Nuclear Physics A} \textbf{\bibinfo{volume}{481}},
  \bibinfo{pages}{81 } (\bibinfo{year}{1988}), 
  \urlprefix\url{http://www.sciencedirect.com/science/article/pii/037594748890%
4745}.

\bibitem[{\citenamefont{Wu and Yang}(1976)}]{wu1976}
\bibinfo{author}{\bibfnamefont{T.~T.} \bibnamefont{Wu}} \bibnamefont{and}
  \bibinfo{author}{\bibfnamefont{C.~N.} \bibnamefont{Yang}},
  \bibinfo{journal}{Nuclear Physics B} \textbf{\bibinfo{volume}{107}},
  \bibinfo{pages}{365 } (\bibinfo{year}{1976}), 
  \urlprefix\url{http://www.sciencedirect.com/science/article/pii/055032137690%
1437}.

\end{thebibliography}

\end{document}